\documentclass[aps,pre,onecolumn,showpacs,superscriptaddress,noeprint,10pt,nofootinbib]{revtex4-1}

\usepackage{amsmath,amssymb,amsfonts}
\usepackage[utf8]{inputenc}
\usepackage[T1]{fontenc}
\usepackage[english]{babel}
\usepackage{graphicx}
\usepackage{tikz}
\usepackage{quantikz}
\usepackage{algorithm}
\usepackage{algorithmic}
\usepackage{xcolor}
\usepackage{soul}           
\usepackage{enumitem}
\usepackage{booktabs}
\usepackage{hyperref} 
\usepackage[normalem]{ulem}
\usepackage{booktabs}

\begin{document}
\title{Variational and Annealing-Based Approaches to Quantum Combinatorial Optimization}

\author{Hala Hawashin}
\email{h.hawashin@exp.science}

\author{Deep Nath}
\email{d.nath@exp.science}

\author{Marco Alberto Javarone}
\email{m.javarone@exp.science}

 \affiliation{Exponential Science Foundation, Lugano, Switzerland}

\begin{abstract}
In this work, we review quantum approaches to combinatorial optimization, with the aim of bridging theoretical developments and industrial relevance. We first survey the main families of quantum algorithms, including Quantum Annealing, the Quantum Approximate Optimization Algorithm (QAOA), Quantum Reinforcement Learning (QRL), and Quantum Generative Modeling (QGM).
We then examine the problem classes where quantum technologies currently show evidence of quantum advantage, drawing on established benchmarking initiatives such as QOBLIB, QUARK, QASMBench, and QED-C. These problem classes are subsequently mapped to representative industrial domains, including logistics, finance, and telecommunications.
Our analysis indicates that quantum annealing currently exhibits the highest level of operational maturity, while QAOA shows promising potential on NISQ-era hardware. In contrast, QRL and QGM emerge as longer-term research directions with significant potential for future industrial impact.
%
\end{abstract}

\maketitle
\section{Introduction}

The inherently probabilistic nature of quantum computation aligns closely with optimisation problems, making them a promising domain for demonstrating quantum advantage. Classical optimisation methods often face significant computational challenges when dealing with large-scale datasets, as the search space grows exponentially with problem size. As a result, classical approaches can become increasingly costly and resource-intensive, both in terms of computational effort and energy consumption~\cite{bertsimas2019machine, Arute2019}. For this reason, time complexity is commonly used to assess which computational paradigms may outperform classical methods, particularly for NP-hard and NP-complete problems.
Several optimisation-focused studies have reported potential quantum speed-ups for specific subclasses of NP problems \cite{Shaydulin2024, xujun2025, lu2025, choi2026, montanaro2025}. A growing body of work \cite{zhu2021, perelshtein2022, willow2025} shows that hybrid and quantum-inspired approaches can already exploit quantum properties, such as probabilistic sampling and non-classical search dynamics, without relying on full fault tolerance. These approaches include quantum simulators that emulate quantum processors, variational algorithms combining parametrised quantum circuits with classical optimisers, and quantum annealers based on adiabatic evolution.
Consequently, optimisation has become one of the earliest areas of industrial exploration for quantum technologies, with applications spanning logistics, finance, telecommunications, and energy systems. Despite this progress, a fundamental gap remains between industrial use cases and the abstract optimisation problem classes typically defined in terms of computational complexity. In this work, we identify a shortlist of optimisation problem classes drawn from major benchmarking frameworks \cite{koch2025quantum, Sharma2025comparative, Sawaya2024HamLib}. These problems are then categorised according to their underlying application structure, namely resource distribution and graph-based optimisation, with the aim to establish a clearer link between abstract problem formulations and industry-specific use cases.
Following this perspective, this review synthesises findings from the existing literature to systematically relate these optimisation problem classes to representative industrial applications. To maintain a focus on practical relevance, we shift attention away from oracle-based methods\footnote{Oracle-based methods remain largely theoretical, serving as abstract models in which a black-box function provides information to the quantum algorithm.} and instead examine more implementable approaches, particularly hybrid quantum machine learning techniques such as variational algorithms, as well as quantum annealing.
The remainder of this work is organised as follows: Section~\ref{sec: Background} reviews the fundamental concepts of quantum technologies; Section~\ref{sec:TimeComplexity} defines time complexity; Section~\ref{sec:algorithms} explores emerging hybrid quantum algorithms and approaches; Section~\ref{sec: Benchmarking} examines existing benchmarking methods; and Section~\ref{sec:Industry} establishes the connection to industry.
\section{Quantum Background and Fundamentals}
\label{sec: Background}
\begin{figure}
    \centering
    \includegraphics[width=0.5\linewidth]{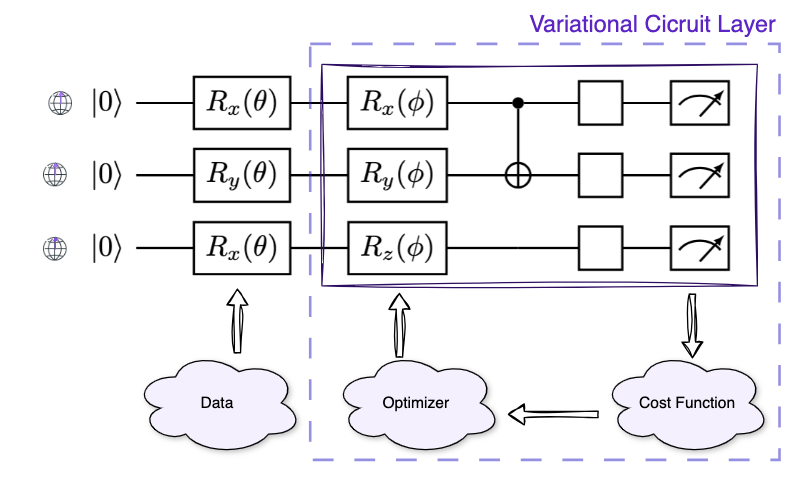}
    \caption{Workflow of variational quantum circuit.}
    \label{fig:PQC}
\end{figure}

Many classical optimisation techniques can be formulated within an energy-based framework, where the objective function is mapped onto a discrete cost landscape, often expressed in binary variables \cite{gladstone2025,han2025,du2022}. In this setting, optimisation proceeds through iterative update rules—deterministic or stochastic—aimed at locating low-energy configurations corresponding to optimal or near-optimal solutions. While effective for moderate problem sizes, such approaches generally face severe scalability limitations as the dimensionality of the search space grows, with performance increasingly constrained by combinatorial complexity and rugged energy landscapes.
Quantum optimisation methods naturally extend this energy-based perspective. Instead of operating directly on classical configurations, quantum systems encode optimisation problems into the spectral properties of an operator, typically a Hamiltonian $H$, whose ground state represents the optimal solution. This formulation induces a structured probability distribution over the solution space, governed by the system’s quantum dynamics rather than explicit enumeration or heuristic search.
Within this framework, quantum evolution—whether adiabatic, variational, or hybrid—biases the probability amplitude toward low-energy configurations, effectively reshaping the exploration of the optimisation landscape. Unlike classical approaches, where transitions between distant local minima may be exponentially suppressed, quantum dynamics enable alternative pathways through the landscape that can modify convergence behaviour for specific problem classes.
The possibility of achieving a computational advantage over classical optimisation methods provides a strong motivation for investigating quantum-based approaches. Early theoretical evidence of such advantage was provided by Shor’s algorithm, which demonstrated exponential speed-up for integer factorisation compared to the best known classical algorithms~\cite{Shor_1997, grover1996, Nielsen_Chuang_2010}. 
Although factorisation is not itself an optimisation problem, this result established that quantum computation can fundamentally outperform classical computation for well-defined computational tasks, thereby motivating the broader exploration of quantum strategies for complex optimisation scenarios.
Quantum computing provides an alternative information-encoding paradigm in which data and optimisation variables are represented within the amplitudes and phases of quantum states in a high-dimensional Hilbert space. For a system of $n$ qubits, the computational state resides in a $2^n$-dimensional space, enabling compact representations of complex optimisation landscapes.
A generic quantum state can be written as $|\psi\rangle = \sum_{i=0}^{2^{n}-1} \alpha_i \, |i\rangle$, where the coefficients $\alpha_i$ encode a probability distribution over computational basis states. In optimisation settings, this representation allows the quantum system to implicitly maintain and evolve a distribution over candidate solutions, rather than operating on individual configurations.
Quantum evolution reshapes this distribution through unitary dynamics, with interference effects favouring low-energy configurations associated with the objective function. In this sense, optimisation can be viewed as a process that progressively biases the probability mass toward regions of the landscape corresponding to near-optimal or optimal solutions.
Entanglement further enriches this representation by introducing non-separable correlations among variables, enabling the exploration of structured dependencies within the search space that are difficult to capture using classical factorised models.
Quantum tunnelling provides a distinctive mechanism for overcoming local minima commonly encountered in classical optimisation. It is a quantum phenomenon in which a particle can penetrate a potential energy barrier even when it lacks the classical energy required to do so~\cite{cohen2019quantum}. Because a qubit is represented by a wave function, a portion of its amplitude can extend through the barrier, resulting in a non-zero probability of appearing on the other side. A common misconception is that exploiting such quantum characteristics necessarily requires a noise-free fault-tolerant quantum computer (FTQC). While this is true for certain oracle-based algorithms, such as Shor’s algorithm or Grover’s search, alternative approaches exist that are compatible with near-term quantum hardware.
Quantum annealers represent some of the earliest quantum devices to demonstrate industrial relevance. They are specialised hardware platforms designed for sampling and optimisation problems. By exploiting adiabatic evolution, annealers gradually transform the system Hamiltonian from an initial form to a problem-specific form, allowing the quantum state to concentrate on low-energy solutions for large-scale tasks. By contrast, gate-based quantum computers operate under a more universal model of computation, currently referred to as noisy intermediate-scale quantum (NISQ) devices. These systems perform computation through discrete sequences of quantum gates that manipulate qubits within a high-dimensional Hilbert space, enabling applications beyond optimisation, including simulation and machine learning. However, this increased generality makes them more susceptible to noise and errors, thereby motivating the development of quantum error correction techniques~\cite{Roffe2019}.
To enable gate-based quantum computers to address industrial optimisation problems under current hardware constraints, most practical use cases rely on hybrid variational architectures, commonly referred to as variational quantum circuits (VQC) or parametrised quantum circuits (PQC). These approaches leverage quantum features such as superposition, entanglement, and quantum encoding, while employing classical optimisation routines to minimise problem-specific cost functions. An ans\"atz\footnote{An ans\"atz is a parametrised trial circuit used to approximate a problem’s solution.} is trained through a feedback loop between quantum and classical processors. The quantum component prepares and measures states that encode the problem structure, while the classical optimiser iteratively updates the circuit parameters to minimise the objective function, as illustrated in Figure~\ref{fig:PQC}. This hybrid architecture is widely regarded as the most promising near-term pathway toward demonstrating practical gate-based quantum advantage.

\section{Time-Complexity and quantum advantage over classical}
\label{sec:TimeComplexity}
Time complexity provides a theoretical framework for analyzing the computational resources required to solve algorithmic problems~\cite{susskind01,susskind02,javarone01,javarone02}.
In particular, time complexity categorises algorithms according to their asymptotic growth rates in a high-level, machine-independent manner, rather than reflecting actual runtime, which may vary depending on hardware architectures, software implementations, and programming languages.
Algorithms are typically classified according to the asymptotic scaling of their runtime with respect to the input size $n$, commonly expressed using Big-O notation. Depending on the functional form of $f(n)$, common examples include linear time $O(n)$, logarithmic or near-linear time $O(n\ln n)$, polynomial time $O(n^k)$, and exponential time $O(k^n)$, where $k$ is a constant. Among these categories, polynomial time is often regarded as the practical threshold separating tractable from intractable problems.
$P$ problems constitute a class of computational problems that are solvable and tractable in polynomial time. Problems that are not solvable in polynomial time using deterministic machines may require exponential time; however, once a solution is found, it can be verified in polynomial time. Such problems belong to the $NP$ (Non-deterministic Polynomial Time) class. $NP$-hard problems represent the most challenging category of problems to solve in polynomial time. A problem $X$ is said to be NP-hard if every problem $Y \in \text{NP}$ is reducible to $X$, such that $Y \leq_p X$. If a problem is both in $NP$ and NP-hard, then it belongs to the $NP$-complete class.
The time complexity of a problem depends on the nature of the algorithm designed to solve it, as well as on the underlying physical and mathematical principles. Classical algorithms generally rely on the foundational concepts of classical physics and computation, and are constrained by principles such as binary logic, locality, and deterministic evolution. By contrast, quantum algorithms exploit quantum features including superposition, entanglement, interference, and probabilistic measurement, potentially providing alternative computational pathways. As a result, the implementation of quantum algorithms can, in certain cases, lead to improvements in time complexity relative to classical approaches.
A prominent example is integer factorisation. The classical general number field sieve (GNFS) algorithm has time complexity
$exp(((c) + o(1))(\log N)^{(1/3)}(\log \log N)^{(2/3)}) \approx exp(O(n^{1/3}\log{n}^{1/3}))$ for $N \to \infty$, where $c=(64/9)^{1/3}=1.9223$ and $n=\log{N}$, corresponding to sub-exponential scaling. In contrast, Shor’s algorithm exploits quantum mechanics to reduce integer factorisation to a problem solvable in polynomial time on a quantum computer, thereby achieving an exponential speed-up.
Another illustrative example is Grover’s algorithm, in which the time complexity of unstructured search is reduced from $O(N)$ to $O(\sqrt{N})$ by leveraging quantum amplitude amplification. To further demonstrate the potential advantages of quantum algorithms over classical counterparts, several additional cases have been examined in~\cite{vaezi2023quantum}. Nevertheless, quantum algorithms do not universally outperform classical ones. Their effectiveness depends critically on algorithmic structure, the quantum resources employed, and the extent to which these features are well matched to the problem under consideration.

\begin{figure}
    \centering
    \includegraphics[width=0.5\linewidth]{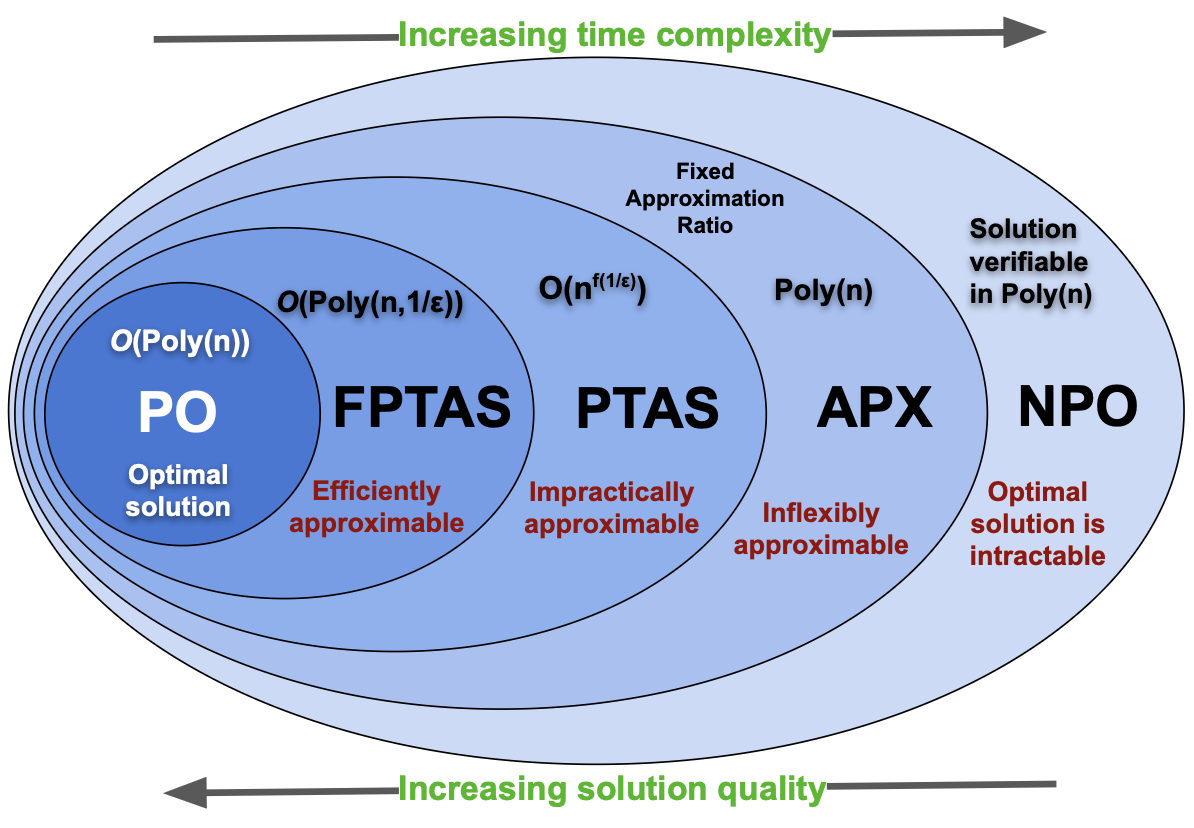}
    \caption{Time complexity, solution quality associated with different complexity classes of optimization problems. Here, $O(Poly(n))$ signifies polynomial time complexity in input size, $n$.}
    \label{fig:TC}
\end{figure}
In optimisation theory, the complexity classes discussed above admit precise analogues. In particular, PO and NPO denote the classes of optimisation problems corresponding to the classical decision classes $P$ and $NP$, respectively.
Several important subclasses of NPO are defined according to their approximability properties, including APX (Approximable), PTAS (Polynomial-Time Approximation Scheme), and FPTAS (Fully Polynomial-Time Approximation Scheme). Problems in APX admit polynomial-time approximation algorithms with a fixed approximation ratio bounded by a constant $c$.
Among other subclasses of NPO, PTAS and FPTAS problems allow polynomial-time algorithms to compute solutions within a factor of $1+\epsilon$ of the optimal value, where $\epsilon > 0$. However, an FPTAS additionally requires that the runtime be polynomial in both the input size $n$ and the approximation parameter $\epsilon$.
By contrast, while the runtime of a PTAS is polynomial in $n$, it may be exponential in $1/\epsilon$. As a consequence, approximate solutions for problems admitting an FPTAS can typically be computed more efficiently than those requiring a PTAS. Figure~\ref{fig:TC} provides a schematic illustration of the optimisation-related complexity classes discussed in this section.
\section{Quantum Methods for Optimization-based Problems}
\label{sec:algorithms}
%

Quantum optimisation is realised on two hardware paradigms: analogue annealing systems and gate-based universal processors. Annealers encode optimisation problems directly into Ising or QUBO Hamiltonians \cite{deFalco2011, Hauke_2020}, with performance guided by dynamical evolution toward low-energy configurations. This native embedding makes annealing architectures particularly effective for combinatorial optimisation and probabilistic sampling. 
Gate-based processors, by contrast, offer a more universal approach under the DiVincenzo framework \cite{DiVincenzo_2000}, enabling optimisation through circuit-based formulations such as variational ans\"atze. For practical deployment, research on gate-based optimisation has focused predominantly on NISQ-compatible algorithms \cite{Preskill2018}, as fault-tolerant architectures remain beyond current hardware capabilities.

%
Algorithms designed for execution on fault tolerant quantum computers (FTQCs) are typically difficult to deploy on near-term hardware and often assume noise-free operation, rendering them impractical for current industrial applications. Accordingly, this work focuses on near-term, hardware-compatible approaches to optimisation problems, grouped according to their learning objectives, and excludes methods that explicitly require fault tolerance. Nevertheless, research on FTQC-compatible algorithms remains active, particularly in the context of optimisation, including quantum walks~\cite{Campos_2023}, quantum Gibbs sampling~\cite{basso2024}, Grover’s search~\cite{nagy2024}, and quantum Markov Chain Monte Carlo methods~\cite{ozgul2025}.

\begin{figure}[H]
    \centering
    \includegraphics[width=0.8\linewidth]{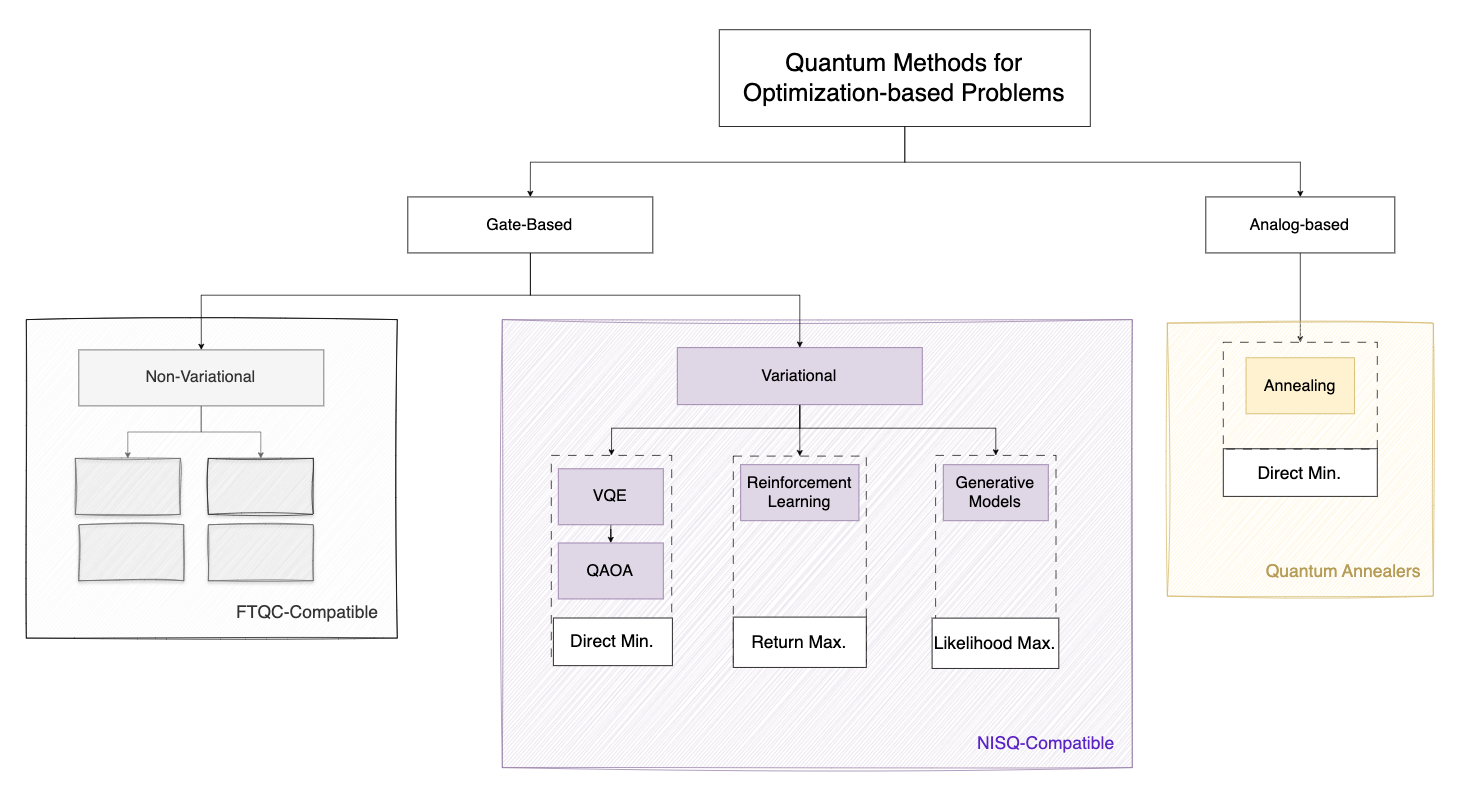}
    \label{fig:placeholder}
    \caption{The hierarchical diagram outlines the relationships between quantum optimization methods, which are broadly divided into gate-based and analogue-based approaches. Gate-based methods include fault-tolerant quantum computing (FTQC), shown in grey and excluded as out of scope, and NISQ approaches, which are more practical given current hardware constraints. Analogue-based methods primarily encompass quantum annealing. At the base of each branch is the corresponding objective function adopted by the method. These approaches are categorized by their optimization goals: minimizing a cost function (direct cost minimization), maximizing expected outcomes (return maximization), or increasing the likelihood of observed data (likelihood maximization).}
\end{figure}

\subsection{Direct Cost Minimisation}
Direct cost minimisation represents the most straightforward approach to optimisation problems~\cite{Hadfield2019}. It targets the problem objective directly, without relying on surrogate rewards, adversarial formulations, or policy-based return maximisation. The optimisation problem is first encoded as a QUBO or an equivalent Ising Hamiltonian, after which the solver searches for the ground-state energy configuration.
In practice, this is achieved either (i) through a quantum annealer, which performs a derivative-free search for low-energy states via adiabatic evolution, or (ii) through a variational, gate-based quantum circuit whose parameters are iteratively tuned by a classical optimiser to minimise the expected energy of the Hamiltonian.
\subsubsection{Annealing}
\begin{figure}[H]
    \centering
    \includegraphics[width=0.6\linewidth]{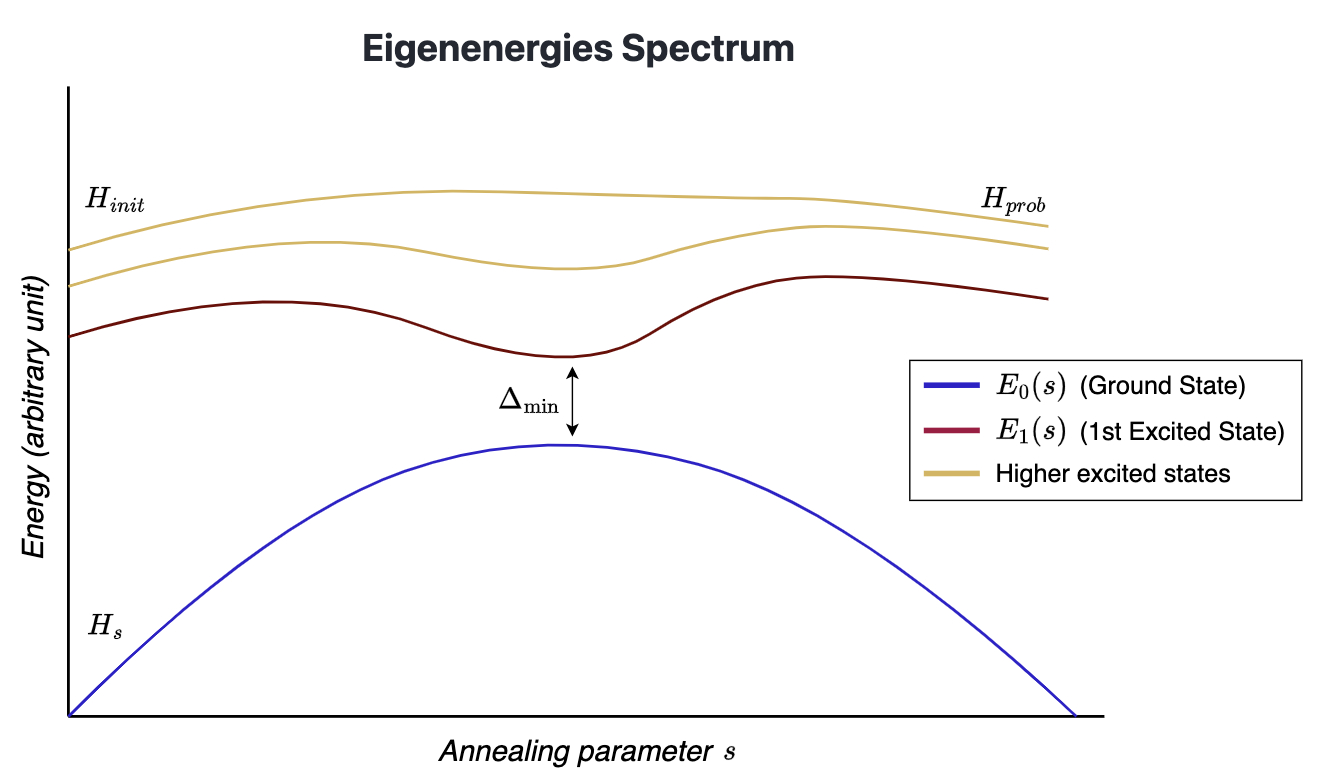}
    \caption{Eigen-energy spectrum illustrating quantum annealing dynamics.  The ground and excited state energies $E_0(s)$ and $E_1(s)$ evolve as the annealing parameter $s$ varies.  The minimum gap $\Delta_{\min}$ marks the point of closest approach between the two lowest eigenstates,  which constrains the required annealing rate to maintain adiabatic evolution.}
    \label{fig:AnnealingDiag}
\end{figure}

Quantum annealing formulates discrete optimisation as as an energy-optimisation problem. In contrast to circuit-based approaches that require explicit gate decompositions, quantum annealing encodes the problem directly into a continuous physical evolution via a time-dependent Hamiltonian. The hardware then interpolates between an initial and problem Hamiltonian, reshaping the energy landscape so that the quantum state evolves toward low-energy configurations corresponding to candidate optima.
The cost function is typically expressed in either Ising or QUBO formalism.
The Ising model, originating in statistical mechanics, defines the energy of a spin configuration as
\[
 E(s)
= \underbrace{\sum_{i=1}^n h_i\, s_i}_{\text{fields}}
\;+\;
\underbrace{\sum_{i=1}^{N-1}\sum_{j=i+1}^N J_{ij}\, s_i s_j}_{\text{couplers}}, \qquad s_i\in\{-1,+1\}.
\]
Whereas the QUBO model, widely used in combinatorial optimisation, defines the energy of a binary configuration as
\[
f(x)
= \underbrace{ \sum_{i} Q_{i,i}\,x_i  }_{\text{fields}}
\;+
\underbrace{\sum_{i<j} Q_{i,j}\,x_i x_j}_{\text{couplers}}, \qquad x_i\in\{0,1\}.
\]
These two formulations are equivalent under a simple spin–binary variable transformation:
\[
s_i = 2x_i - 1 
\quad \Longleftrightarrow \quad 
x_i = \frac{1+s_i}{2}, \quad x_i \in \{0,1\}, \; s_i \in \{-1,+1\}.
\]
Problems are embedded on annealers via an Ising (or equivalent QUBO) Hamiltonian~\cite{Lucas_2014}. The coupling terms govern whether configurations favour alignment or anti-alignment, while hard constraints are incorporated directly through penalties to the same coefficients. As a result, constraint violations raise the system’s energy, biasing the annealing dynamics toward low-energy, feasible configurations.
%
The annealer is initialised in the ground state of a driver Hamiltonian, which corresponds (in the computational basis) to an equal superposition of all bit strings. During the anneal, the driver Hamiltonian $H_{\text{init}}$ is gradually turned off while the problem Hamiltonian $H_{\text{prob}}$ is turned on, following
\[
H(s) \;=\; A(s)\,H_{\text{init}} \;+\; B(s)\,H_{\text{prob}}, \qquad s\in[0,1].
\]
If the interpolation from $H_{\text{init}}$ to $H_{\text{prob}}$ proceeds sufficiently slowly, the adiabatic theorem ensures that a system initially prepared in the ground state of $H_{\text{init}}$ will remain, or remain close to, the instantaneous ground state of the time-dependent Hamiltonian $H(s)$ throughout the evolution~\cite{Farhi_2001}. Under such conditions, the system may end the anneal in the ground state of $H_{\text{prob}}$, thereby yielding the optimal solution to the problem—see Fig.~\ref{fig:AnnealingDiag}.

%
In practice, finite anneal times and non-zero temperature effects limit adiabatic performance, increasing unexpected jumps to excited states. To mitigate these effects, quantum annealers are typically executed multiple times to generate low-energy samples, followed by classical post-processing to identify high-quality solutions~\cite{Albash2018, Zick2015,ayanzadeh2021}.
%
%
\subsubsection{QAOA}
The Quantum Approximate Optimization Algorithm (QAOA)~\cite{farhi2014} can be viewed as a Trotterized, circuit-based approximation to adiabatic optimisation. The classical objective is encoded in a cost Hamiltonian $H_C$, while optimisation is implemented through a parametrised circuit consisting of $p$ alternating applications of the cost operator and a mixer Hamiltonian $H_M$. The mixer prevents the evolution from remaining confined to eigenstates of $H_C$, enabling exploration of the Hilbert space during the optimisation process.

%
\begin{center}
\begin{quantikz}[row sep=0.2cm, column sep=0.3cm]
\lstick{$\ket{0}$} & \gate[wires=3]{H^{\otimes n}} &
            \gate[wires=3]{e^{-i\gamma_1 H_C}} & \gate[wires=3]{e^{-i\beta_1 H_M}} & \push{\cdots} &
            \gate[wires=3]{e^{-i\gamma_p H_C}} & \gate[wires=3]{e^{-i\beta_p H_M}} & \meter{} \\
\lstick{$\ket{0}$} & \qw & \qw & \qw & \push{\cdots}  &  \qw & \qw & \meter{} \\
\lstick{$\ket{0}$} & \qw & \qw & \qw & \push{\cdots}  & \qw & \qw & \meter{}
\end{quantikz}
\end{center}
\paragraph{\boldmath $H_C(\gamma)$ and $H_M(\beta)$.}
The cost Hamiltonian $H_C$ encodes the optimization problem by assigning higher energies to undesirable solutions and the lowest energy (ground state) to the optimal configuration. Its associated unitary, $e^{-i\gamma H_C}$, applies phase shifts proportional to the cost of each computational basis state, thereby biasing the quantum state toward lower-cost solutions. This mechanism plays a role analogous to the couplers in quantum annealing, where interactions encode the objective function.
The mixer Hamiltonian $H_M$, typically a sum of Pauli-$X$ operators, is applied via $e^{-i\beta H_M}$ to induce transitions between computational states, mitigating confinement in local minima. Alternating applications of the cost and mixer unitaries balance exploration (via $H_M$) and exploitation  (via $H_C$),, with increasing circuit depth $p$ enhancing the expressive capacity of the ansatz.
\paragraph{\boldmath $p$-layers}
In QAOA, a circuit of depth $p$ contains $2p$ variational parameters, corresponding to the cost $(\gamma_i)$ and mixer $(\beta_i)$ unitaries at each layer. For example, at $p=1$ the classical optimizer selects two parameters $(\gamma_1, \beta_1)$; at $p=2$, four parameters $(\gamma_1, \beta_1, \gamma_2, \beta_2)$ are optimized, and so on for increasing values of $p$. In the limit $p \to \infty$, QAOA approaches adiabatic quantum computing and, under ideal conditions, converges to the exact ground state, yielding the optimal solution.
Already at shallow depth ($p=1$), QAOA admits provable approximation for certain structured problems, such as MaxCut on regular graphs. In Ref. \cite{FarhiGamrnik2020}, the algorithm was analyzed for structured instances including Max-3-Lin-2, showing that even a single-layer circuit can outperform specific classical heuristics. Notably, this result relied on an analytical determination of optimal parameters, rather than variational learning.
Subsequent studies have highlighted the dependence of solution quality on circuit depth across different application domains. In portfolio optimization, Zaman et al.~\cite{Zaman2024} reported comparable performance at moderate depths ($p=3,5$) on small instances implemented on IBM-Q quantum processors. Brandhofer et al.~\cite{Brandhofer_2022} further observed that, in noiseless simulations, performance continues to improve up to $p=7$, whereas on NISQ hardware performance gains tend to plateau beyond $p=3$–$4$ due to noise accumulation.
In scheduling problems, \cite{KUROWSKI2023518} showed that deeper circuits ($p=6,9$), combined with interpolation-based parameter initialization strategies, significantly increase the probability of obtaining high-quality solutions (up to 90\%). For vehicle routing problems, \cite{azfar2025} demonstrated a working implementation at $p=2$ on IBM Quantum hardware, achieving feasible solutions on NISQ devices. In contrast, Azad et al.~\cite{Azad2023}, using noiseless classical simulations, explored deeper circuits and reported correct solutions with high probability at $p=12$, as well as distributions of feasible solutions at $p=24$. Overall, noiseless simulations consistently indicate that increasing circuit depth $p$ enhances solution quality. On real hardware, however, noise imposes practical limits on achievable depth. Consequently, an alternative and complementary strategy consists in improving parameter initialization and optimization techniques for the $2p$ variational parameters, rather than relying solely on deeper circuits.
\paragraph{\boldmath $2p$ parameters: $(\gamma_i,\beta_i)$}
Given the hardware-imposed limitations on circuit depth $p$, significant attention has shifted toward the efficient optimization of the $2p$ variational parameters $(\gamma_1, \beta_1, \ldots, \gamma_p, \beta_p)$. Direct optimization from random initialization at increasing depths rapidly becomes computationally expensive and may suffer from instability and poor convergence.
To mitigate these issues, interpolation strategies — also referred to as parameter transfer — have been proposed as warm-start techniques \cite{Zhou2020}. The central idea is to initialize the parameters of a circuit of depth $p+1$ using the optimized parameters obtained at depth $p$. When increasing the depth, the optimized angles from the shallower circuit are effectively “stretched” to populate the deeper ansatz. Rather than starting from random guesses, the new parameters are inferred by following the smooth trends observed in the previously optimized values, thereby improving convergence and reducing the optimization overhead.
In the same work, the authors introduced a complementary strategy known as \textit{Fourier parametrization}. Instead of treating each of the $2p$ angles as independent variables, the sequence of QAOA parameters across layers is expressed as a truncated Fourier series. In this representation, the angles are generated from a small number of Fourier coefficients defining sine and cosine modes over the circuit depth. This reduces the effective dimensionality of the optimization landscape and often improves trainability.
A parallel research direction has explored learning-based predictors for QAOA parameter selection. Rather than relying exclusively on classical optimizers during each execution, machine learning models are trained offline on smaller instances, where optimal or near-optimal parameters can be computed. Approaches include supervised regression models \cite{alam2020}, neural network architectures \cite{Xie_2023}, and reinforcement learning agents \cite{Wauters_2020}, which aim to generalize parameter schedules to larger or previously unseen instances. Additional methods leverage clustering of parameter landscapes or meta-learning strategies to transfer information across related problem classes \cite{Sureshbabu_2024} \cite{galda2023}. By shifting part of the computational effort to an offline training phase, these approaches reduce the number of expensive quantum circuit evaluations required during optimization, thereby enhancing scalability.
However, increasing the flexibility of parameterisation can also enlarge the effective search space, potentially leading to exploration of infeasible or poorly structured regions that undermine optimization efficiency. To address this, \cite{Wang2023} proposed an enhanced architecture termed Quantum Alternating Operator Ansatz+, which incorporates a constraint-aware mixer. In this formulation, problem-specific constraints are embedded directly into the quantum evolution, ensuring that the state remains within the feasible subspace throughout the circuit. This restriction reduces the search space and improves solution quality for constrained optimization problems.
Other machine learning-based approaches similarly aim to reduce classical optimizer evaluations by predicting the $2p$ parameters. Supervised regression models have been employed to approximate optimal schedules \cite{alam2020}, while reinforcement learning controllers have been trained to select $(\gamma,\beta)$ sequences that emulate optimal adiabatic performance on benchmark systems such as the Ising chain \cite{Wauters_2020}, and to learn update strategies that generalize across varying problem sizes \cite{khairy2019}.

\subsection{Return Maximisation}
Reinforcement learning considers optimisation of a policy $\pi$ through interaction with an environment, where actions are selected based on the current state $s$ to maximise expected return. Approaches broadly divide into model-based methods, which learn or approximate the transition dynamics $P(s'|s,a)$ for planning, and model-free methods, which update policies or value estimates directly from sampled interaction data~\cite{sutton2018}.  Algorithmically, this can be value-based, policy-based, or a hybrid approach. In value-based methods, the agent estimates a value function, such as the action-value function $Q(s,a)$, and selects actions by comparing these estimates. In policy-based methods, the agent learns a parametrized policy $\pi_\theta(a|s)$ directly, adjusting the parameters $\theta$ through gradient ascent on the expected return. Actor–critic methods combine both perspectives: the actor selects actions according to a policy, while the critic evaluates them by estimating a value function, thereby stabilizing learning.
\begin{figure}
\centering
\includegraphics[width=0.5\linewidth]{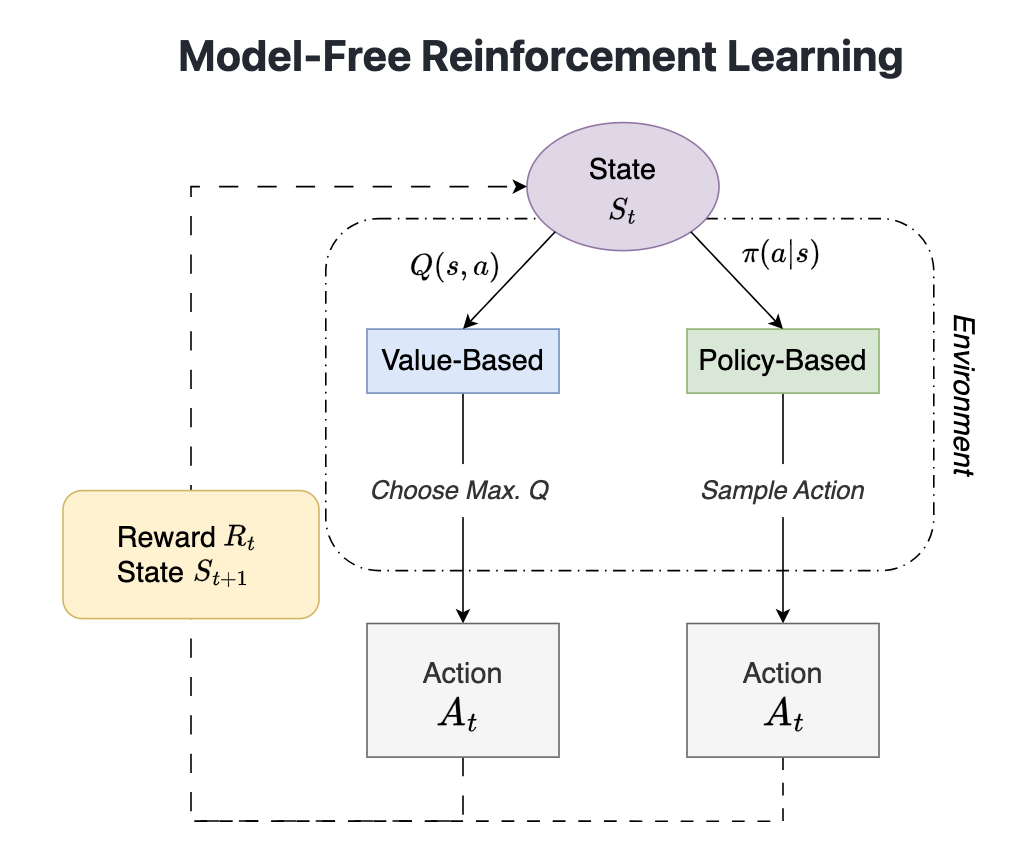}
\caption{Schematic representation of the agent–environment interaction in reinforcement learning.}
\label{fig:placeholder}
\end{figure}
The quantum extension of reinforcement learning (QRL) typically employs Variational Quantum Circuits (VQCs). Measurements of a VQC naturally produce probability distributions over actions, making policy representations straightforward. However, policy-gradient methods suffer from high-variance estimates, a limitation amplified on NISQ hardware by measurement noise and finite sampling. Value-based approaches face a different challenge, as accurate value estimation requires extensive circuit evaluations, leading to significant sampling overhead. Consequently, hybrid actor--critic architectures are often adopted, where a VQC implements the policy (actor) while a classical critic provides lower-variance value estimates to stabilise training.

When QRL is applied to combinatorial optimization, the structure of the problem differs from standard sequential decision-making. The search space is discrete, and each candidate solution is associated with a single objective value. There is no extended reward trajectory to estimate; only the quality of the final solution matters. Consequently, policy-search approaches are often more natural than value-based methods in this context, as discussed in \cite{khairy2019}.

Kruse et al. \cite{kruse2024} explicitly addressed this perspective when comparing QRL with QAOA. They proposed a Hamiltonian-based QRL framework tailored to combinatorial optimization, where the QUBO formulation is embedded directly into the ans"atz of the QRL agent. In this design, the learning process is guided by the problem Hamiltonian itself, restricting exploration to a problem-informed search space. Their results showed that, despite requiring a larger number of circuit evaluations, the Hamiltonian-informed QRL approach outperformed QAOA in terms of solution quality and generalization to unseen instances. Furthermore, feasibility was enforced by masking invalid solutions, ensuring that the agent consistently produced valid outputs.

A different approach was proposed in \cite{sequeira2023}, where a policy-gradient QRL method based on a VQC was introduced without explicitly encoding the problem Hamiltonian into the circuit structure. Instead, the model learns directly from measurement outcomes: each circuit execution produces a candidate solution, which is evaluated using the objective function. This formulation maps naturally onto combinatorial optimization, where the agent samples discrete configurations and progressively shifts probability mass toward higher-quality solutions, in a manner analogous to probabilistic exploration in QUBO and Ising formulations.
\subsection{Likelihood Maximisation}
Unlike direct cost minimization or return maximization, likelihood maximisation directly reshapes the model’s probability distribution so that high-probability outcomes coincide with desirable or observed configurations.
Generative models aim to approximate the probability distribution underlying a dataset. In practice, they rely on a parametrised generator that produces samples from a latent or prior distribution, together with a feedback signal that evaluates sample quality with respect to a target objective. Model parameters are updated iteratively so that the generated distribution progressively aligns with the target data distribution.
Although numerous generative modelling paradigms exist, only a subset align naturally with optimisation settings \cite{Romero2017, minami2025}. In particular, energy-based formulations—such as variational, flow-based, and diffusion approaches—frame learning in terms of a scalar energy or cost function \cite{Bond2022}. Under this perspective, optimisation can be viewed as sampling within an energy landscape while progressively concentrating probability mass around low-energy configurations corresponding to candidate optima.
\paragraph{Quantum Variational Models.}
Quantum Circuit Born Machines (QCBMs) and Quantum Variational Autoencoders (Q-VAEs) follow the general structure of variational generative models, employing a quantum or quantum-inspired circuit as the generator and a classical optimizer to tune its parameters. 
Such approaches represent candidate solutions through a quantum state, with optimisation concentrating probability mass around low-cost configurations \cite{Liu2018}. This formulation closely parallels the Variational Quantum Eigensolver (VQE) — also underlying QAOA — and represents one of the most mature and experimentally investigated forms of quantum generative modelling \cite{Liu_2022} \cite{zhang2025}.
\paragraph{Quantum Flow-based Models.}
Flow-based models, including Quantum Normalizing Flows, constitute an emerging research direction. They rely on invertible transformations that map a simple prior distribution into a more complex target distribution aligned with desirable solutions. This structure is conceptually well matched to quantum mechanics, since quantum evolutions are unitary and therefore inherently reversible. Despite this theoretical compatibility, current research on quantum flow models has primarily focused on anomaly detection \cite{rosenhahn2024} and sampling tasks \cite{lawrence2024}, with comparatively limited exploration in combinatorial optimization contexts.
\paragraph{Quantum Diffusion Models.}
Quantum Diffusion Models (QDMs) extend classical diffusion-based generative frameworks to quantum settings by learning to invert a noise-driven evolution process. Like flow-based methods, they typically rely on continuous representations, which can complicate their application to discrete combinatorial optimisation problems. Nonetheless, diffusion-based models have shown strong connections to optimisation in classical settings. Notably, \cite{ma2024} and \cite{zhao2025disco} show how diffusion processes can approximate optimization as progressive energy minimization, a viewpoint that conceptually extends to quantum settings.
More recent approaches propose alternative quantum generative architectures. In \cite{minami2025}, building upon \cite{nakaji2025}, the authors introduce the Conditional Generative Quantum Eigensolver (GQE), a context-aware quantum circuit generator based on an encoder–decoder transformer architecture. In this formulation, combinatorial optimization is treated as a generative task, where optimal circuit configurations are produced conditioned on specific problem instances. Their results indicate that GQE achieves performance comparable to, and in some benchmarks exceeding, QAOA on 10-qubit combinatorial optimization tasks.
A complementary line of research explores quantum-inspired classical architectures that emulate quantum generative behaviour. For example, \cite{Alcazar2024} propose a Generator-Enhanced Optimization (GEO) framework applied to portfolio optimization using real S\&P 500 data. The core component is a Tensor-Network Born Machine acting as a probabilistic generator capable of learning correlations among high-quality candidate solutions. The GEO architecture follows iterative cycles of generation, evaluation, and parameter updates. Owing to the extensive development of tensor-network methods, this approach is generally more mature and scalable than current NISQ-based QAOA implementations. In particular, tensor-network representations such as Matrix Product States can efficiently encode high-dimensional probability distributions while maintaining tractable computational costs.
\section{Benchmarking related to quantum optimization}
\label{sec: Benchmarking}
Systematic benchmarking methodologies are essential for evaluating optimization approaches. They reveal differences in performance, scalability, and efficiency between quantum and classical techniques. In contrast to classical optimization benchmarks, hybrid quantum–classical processes, sampling overhead, device noise, and compilation constraints must be taken into consideration when evaluating performance in quantum computing. Benchmarking is also crucial in reducing the gap between theoretical advancements and real-world implementation in the age of quantum technologies. Three complimentary levels are commonly used to explain benchmarking in quantum optimization. \cite{Rudi2022, koch2025quantum}:

\begin{enumerate}
\item \textbf{Applications Benchmarking}:
The primary objective at this level is to identify the most effective algorithm — classical or quantum — for solving a specific problem instance. These are model-agnostic benchmarks to ensure fair comparison across heterogeneous approaches with respect to real-world tasks.
\item \textbf{Algorithm Benchmarking}:  
Algorithm benchmarking evaluates performance independently of hardware to analyze asymptotic scaling and hyperparameter sensitivity. It identifies architectural bottlenecks and provides the primary metrics for optimizing algorithmic design.
\item \textbf{System Benchmarking}:  
System benchmarking optimizes a fixed algorithm on target hardware for specific problem instances. It accounts for platform-specific constraints, including error mitigation and calibration protocols, effectively providing an application-oriented characterization of hardware performance.
\end{enumerate}
\subsection{Benchmarking Libraries and Problem Sets}
Curated datasets for combinatorial optimization provide standardized instances with controlled scaling properties, enabling reproducible benchmarks across heterogeneous architectures and hybrid optimization pipelines.

\subsubsection{QOPTLib}

QOPTLib is a quantum-computing-oriented benchmark suite for combinatorial optimization problems \cite{Osaba2023qoptlib}. It includes $40$ NP-Hard problem instances drawn from four well-known classes: the Maximum Cut problem (MCP), Bin Packing problem (BPP), Vehicle Routing problem (VRP), and Traveling Salesman problem (TSP). Designed with diverse instance scales, QOPTLib facilitates the study of scaling behaviour and the impact of system size on quantum algorithm performance. In its original study, QOPTLib was used to compare two quantum annealing-based D-Wave solvers: the pure QPU $Advantage_system6.1$ and the hybrid quantum–classical $LeapHybridBQMSampler$. However, the library is hardware-agnostic in principle and can be employed to evaluate both gate-based and quantum annealing approaches, making it a versatile testbed for real-world optimization scenarios.
\subsubsection{QOBLIB}
QOBLIB provides a standardized framework for tracking progress in quantum optimization via the 'Intractable Decathlon' -- a curated set of ten NP-hard instances spanning network design, scheduling, and finance \cite{koch2025quantum}, summarized in Table \ref{tab:ID}. By explicitly integrating hardware-specific constraints with standardized evaluation criteria, the library enables a rigorous, comparative assessment of practical quantum advantage against state-of-the-art classical baselines.
\begin{table*}[h]
\tiny
    \centering
    \caption{The Intractable Decathlon: Condensed Overview}
    \label{tab:ID}
    \begin{tabular}{|p{0.17\textwidth}|p{0.25\textwidth}|p{0.5\textwidth}|}
        \hline
        \textbf{Problem Class} & \textbf{Core Problem} & \textbf{Key Challenges and Characteristics} \\
        \hline
        \textbf{Market Split} & Find a binary vector $x$ satisfying a multi-dimensional subset-sum constraint ($Ax=b$). & NP-hard. Hard for classical solvers to find a feasible solution even for small instances ($\sim$100 variables). \\
        \hline
        \textbf{LABS} (Low Autocorrelation Binary Sequences) & Minimize autocorrelation energy in a binary sequence ($s_j \in \{-1, +1\}$). & Binary structure suitable for Ising models. Classical algorithms struggle to find good solutions for relatively small sizes. \\
        \hline
        \textbf{Minimum Birkhoff Decomposition} & Decompose a doubly stochastic matrix $D$ into the minimum number of permutation matrices ($k$). & NP-hard, difficult to solve exactly even for small instances. Relates to unitaries, fundamental to quantum computing. \\
        \hline
        \textbf{Steiner Tree Packing} & Find a minimum-cost forest packing $n$ node-disjoint Steiner trees into a graph (VLSI routing). & Computationally very hard. Hard for heuristics because one bad decision can immediately lead to infeasibility. \\
        \hline
        \textbf{Sports Tournament Scheduling} & Find a feasible phased double round-robin schedule respecting complex capacity, break, and game constraints. & High combinatorial complexity. Established heuristics often fail to find feasible solutions for medium-sized instances. \\
        \hline
        \textbf{Portfolio Optimization} & Binary quadratic program to balance returns and risk (mean-variance) over multiple time periods with transaction costs. & NP-hard. Challenging for classical stochastic dynamic programming under realistic conditions (e.g., transaction costs). \\
        \hline
        \textbf{Maximum Independent Set (MIS)} & Find the largest set of non-adjacent vertices in an unweighted graph. & A classic strongly NP-hard problem. Converts well to relatively sparse QUBOs with small coefficients. \\
        \hline
        \textbf{Network Design} & Construct a directed graph with fixed in/out-degree $p$ to minimize the maximum edge flow/load. & NP-hard. Trivial to find a feasible solution, but the optimal solution has historically proven elusive even for moderately sized instances. \\
        \hline
        \textbf{Vehicle Routing Problem (VRP)} & Determine cost-minimizing routes for a fleet of $K$ capacitated vehicles serving $n$ customers from a depot. & NP-hard. Classical heuristics are highly optimized and successful, setting a high bar for quantum competition in the near term. \\
        \hline
        \textbf{Topology Design (ODP)} & Minimize the diameter (max shortest path) of a graph given a fixed number of vertices ($n$) and max degree ($d$). & Difficult in practice, with a compact formulation (specified by three integers). Uses binary variables suitable for quantum encoding. \\
        \hline
    \end{tabular}
\end{table*}

\subsubsection{SMU Quantum Optimization Benchmark Library}
Standardized examples of the Market Share, Maximum Independent Set, Multi-Dimensional Knapsack, and Quadratic Assignment tasks are offered by the SMU Quantum Optimization Benchmark \cite{Sharma2025comparative}. By comparing quantum and hybrid heuristics to the most advanced classical baselines in formulations that are important to the industry, this library makes cross-platform scalability analysis easier.
\subsubsection{HamLib}
HamLib offers an extensive collection of many-body and optimization problems that are directly mapped to qubit Hamiltonians \cite{Sawaya2024HamLib}. Its native representation enables benchmarking for variational heuristics and quantum annealing in both application-driven and physics-inspired domains, spanning combinatorial problems (e.g., Max-k-SAT) and physically motivated models (e.g., Fermi–Hubbard and Heisenberg).

\subsubsection{Other Available Benchmarking Libraries}

In addition to quantum-focused repositories, several well-established classical benchmark libraries are routinely employed by quantum researchers to assess performance against classical baselines. These include TSPLIB (Traveling Salesman Problem Library) \cite{Reinelt1991TSPLIB}, QAPLIB (Quadratic Assignment Problem Library) \cite{Burkard1997QAPLIB}, MIPLIB (Mixed Integer Programming Library) \cite{Gleixner2021MIPLIB2017}, and the OR-Library \cite{Beasley1990ORLibrary}. These datasets provide standardized, extensively studied instances that enable rigorous cross-paradigm comparisons.

In machine learning, MNIST contains 60,000 training images and 10,000 testing images of handwritten digits (0–9), for a total of 70,000 grayscale samples \cite{LeCun1998MNIST, LeCun2010MNIST}. While primarily designed for classical deep learning, MNIST has also been widely adopted in quantum machine learning research as a proof-of-concept dataset.

This ecosystem is further expanded by specialized resources such as QDataSet and PennyLane Quantum Datasets, which offer curated, pre-processed data for quantum control, tomography, and chemistry, guaranteeing reproducible evaluation across both optimization and learning paradigms \cite{Perrier2022QDataSet, Bergholm2018PennyLane}.
\subsection{Advanced Benchmarking Frameworks}

Our main motivation in this review is to talk about application-centric benchmarking methods, particularly when it comes to optimization challenges. For industry-related applications like hardware design and problem solving, the following benchmarking frameworks could be useful.

\subsubsection{QASMBench}

A benchmarking suite called QASMBench \cite{Ang2023} was created to assess the performance of quantum compilers, classical quantum simulators, and NISQ devices \cite{Preskill2018}. This is a combination of quantum programs written in Open Quantum Assembly Language (OpenQASM) \cite{cross2017}, which is a unified low-level assembly language for quantum machines. This suite comprises a wide variety of quantum problems, spanning from chemistry and optimization to physics and error correction, as well as cryptography and machine learning. Depending on how many qubits are needed to solve the underlying problem, the quantum problems are divided into three categories here: small-scale (2–5 qubits), medium-scale (6–15 qubits), and large-scale (>15 qubits). 
QASMBench also introduces several metrics to to analyze structural, temporal, and operational characteristics of quantum circuits. Table~\ref{tab:metrics} summarizes these metrics. They describe different properties of quantum circuits and help explain how circuits stress quantum hardware. Structural and temporal metrics describe the basic resources needed to run a circuit. Utilization and interface metrics explain how measurements interact with conventional processing and how well the hardware is used. The performance of quantum circuits using NISQ devices can be impacted by hardware constraints like coherence time and qubit connection, which are reflected in coherence and topological metrics.
\begin{table}[h]
\centering
\small
\caption{Architectural and operational metrics used in QASMBench for evaluating NISQ hardware performance.}
\label{tab:metrics}
\begin{tabular}{|p{0.20\textwidth}|p{0.10\textwidth}|p{0.60\textwidth}|}
\hline
\textbf{Metric} & \textbf{Category} & \textbf{Analytical Focus and Significance} \\
\hline
Circuit Width & Structural & Number of active qubits involved in superposition during circuit execution. Indicates hardware resource requirements. \\
\hline
Circuit Depth & Temporal & Minimum number of sequential gate layers required to execute the circuit. Reflects total execution time. \\
\hline
Gate Density & Utilization & Measures the degree to which available qubit slots are occupied by gate operations during execution. \\
\hline
Measurement Density & Interface & Evaluates measurement overhead and interaction between quantum and classical processing. \\
\hline
Retention Lifespan & Coherence & Quantifies how long qubits remain active relative to hardware coherence limits ($T_1/T_2$). \\
\hline
Entanglement Variance & Topological & Measures imbalance in two-qubit gate distribution and identifies possible ``qubit hotspots''. \\
\hline
\end{tabular}
\end{table}
When benchmarking hardware performance, fidelity is often used as the primary evaluation metric:

\begin{equation}
F(\rho, \sigma) =
\left(Tr\sqrt{\sqrt{\rho}\sigma\sqrt{\rho}}\right)^2
\end{equation}
\noindent where $\rho$ and $\sigma$ denote the density matrices corresponding to the ideal and noisy quantum states, respectively \cite{Ang2023}. By combining circuit-level metrics with fidelity-based evaluation, QASMBench provides an application-oriented framework for assessing the performance of NISQ hardware platforms such as IBM-Q, Rigetti, IonQ, and Quantinuum. Hardware-level trade-offs between coherence time, gate quality, and device connectivity have a significant impact on near-term quantum optimization algorithms' practical performance. Experimental investigations using the QASMBench platform show that hardware capability indicators such as Quantum Volume are not always directly related to execution fidelity for true quantum circuits. Rather, circuit-level features such as depth, width, and entanglement structure often influence NISQ device performance. Cross-platform evaluations, where the possible circuit quality is dictated by the interaction of coherence time, gate fidelity, and connection constraints, further highlight the architectural differences between trapped-ion and superconducting processors. These results suggest that the successful implementation of quantum optimization algorithms will depend on hardware-aware circuit compilation, qubit mapping strategies, and topology-aware optimization techniques.

\subsubsection{QUARK: QUantum computing Application benchmaRK}

Low-level quantum benchmarks often focus on hardware properties such as gate fidelity or coherence time. However, these metrics do not necessarily reflect the performance of quantum algorithms on real-world problems. To address this limitation, application-level benchmarking frameworks have been proposed. One such framework is QUARK (QUantum computing Application benchmaRK), which enables systematic evaluation of end-to-end quantum applications \cite{Rudi2022}.

QUARK concurrently handles five different benchmarking layers: problem definition, problem translation into mathematical formulation, feasible solution finding, hardware-specific task submission, and benchmarking management \cite{Rudi2022}. The key metrics associated with the benchmarking techniques are:

\begin{enumerate}
    \item \textbf{Time-to-Solution ($TTS$):} end-to-end time required to obtain the solution. Mathematically,
    \begin{equation}
        TTS=T_{mapping}+T_{solver}+T_{reverseMap}+T_{processSolution}+T_{validation}+T_{evaluation}.
    \end{equation}
    Here, $T_{mapping}$, $T_{solver}$, $T_{reverseMap}$, $T_{processSolution}$, $T_{validation}$ and $T_{evaluation}$ are times associated to the mapping of problem to mathematical formulation, actual algorithm execution, converting solution back to application domain, post-processing steps, checking constraint satisfaction and computing quality metrics.
    \item \textbf{Validity ($V$):} examines if the solution produced by a specific solver is valid by taking into account all the constraints. $V$ is application-specific, and the corresponding mathematical function is customizable.
    \item \textbf{Quality ($Q$):} measures the solution quality within the context of real-world industrial workloads.
\end{enumerate}
QUARK is particularly relevant for benchmarking quantum optimization algorithms. Many optimization problems are first formulated as combinatorial problems and then mapped to mathematical models such as QUBO or Ising Hamiltonians. These formulations can be solved using different approaches, including QA, QAOA, variational quantum algorithms, and classical optimization methods. QUARK makes it possible to compare these methods systematically across diverse optimization problem cases by supporting a variety of mappings, solvers, and hardware backends. This enables research into the effects of problem size, formulation, and hardware limitations on algorithm performance.
QUARK offers a repeatable paradigm for assessing the effectiveness of quantum optimization techniques on classical, quantum, and hybrid computer systems by combining the metrics $TTS$, $V$, and $Q$. The objective of employing these three interrelated measures is to offer a thorough, consistent, and repeatable assessment of application performance that surpasses fundamental hardware metrics. Additionally, QUARK analyses different infrastructures (such as D-Wave quantum annealers, simulators, and classical solvers) and explores the potential opportunities for quantum solvers to address real-world problems. Quark has also been effectively used in the context of quantum machine learning (QML) methods \cite{kiwit2024benchmarking}, particularly the quantum generative adversarial network (QGAN) \cite{benedetti2019generative} and the quantum circuit Born machine (QCBM) \cite{lloyd2018quantum}.

\subsubsection{Advanced QED-C: Operational Utility and Performance Trade-offs}

In recent years, researchers have become increasingly interested in the need for appropriate application-focused benchmarking methods that go beyond basic fidelity and gate-count measurements. A significant method is the Quantum Economic Development Consortium (QED-C), which provides a flexible framework for testing the performance of quantum computing systems, including both gate-based and annealing hardware \cite{sri_international_2020_applicationoriented}. However, a relatively new and advanced version of QED-C places a strong emphasis on operational performance matrices \cite{lubinski2024optimization}. The advanced QED-C has been shown to be a valuable tool to establish a connection between execution time and the quality of the solution associated with specific real-world problems. These challenges, which include network design, portfolio selection, and logistics routing, require high-quality results within a specific time-frame. Therefore, comprehending the trade-off between the quality (quality metrics) and the speed (time metrics) of a solution is essential to understanding the importance of quantum algorithms and technologies for real-world combinatorial optimization problems.

\begin{enumerate}

\item{\textbf{Solution quality metrics:}}

To capture the statistical distribution of optimization outcomes, the framework employs several complementary quality metrics rather than relying solely on expectation values:

\begin{itemize}

\item \textbf{Approximation Ratio ($AR$):}  
Normalizes the expected energy of the quantum state $\langle E \rangle$ relative to the optimal ground-state energy $E_{\min}$, where

\[
AR = \frac{\langle E \rangle}{E_{\min}} .
\]

\item \textbf{Conditional Value at Risk ($CVaR_{\alpha}$):}  
A tail-performance metric that averages only the lowest-energy $\alpha$ fraction of sampled outcomes. This metric emphasizes the best candidate solutions and is therefore particularly useful for evaluating heuristic optimization algorithms.

\item \textbf{Gibbs Objective:}  
An exponentially weighted performance metric in which a tunable parameter $\eta$ controls the emphasis placed on low-energy solutions, enabling consistent comparison across different optimization algorithms and solver architectures.

These quality metrics -- expectation-based (AR), tail-based (CVaR), and exponentially-weighted (Gibbs) -- offer a multifaceted perspective on the performance of algorithms.

\end{itemize}

\item{\textbf{Temporal Metrics}}

A key component of the framework is the decomposition of the total elapsed execution time

\[
t_{\text{elapsed}} =
t_{\text{queue}} +
t_{\text{compile}} +
t_{\text{load}} +
t_{\text{quantum}} +
t_{\text{classical}},
\]

which separates quantum execution time ($t_{\text{quantum}}$) from classical overheads such as compilation, queue delays, and parameter optimization. This decomposition enables identification of performance bottlenecks within hybrid quantum–classical workflows. We can also have, two standard OR metrics are derived:
\begin{itemize}
    \item \textbf{Time-to-Target (TTT):} Expected time required to reach a desired quality threshold.
    \item \textbf{Time-to-Solution (TTS):} Expected time required to find the optimal solution.
\end{itemize}
These metrics are especially important for time-constrained optimization tasks such as real-time resource allocation or market response modeling.

\item \textbf{Visualization Framework: The Area Plot}

A notable innovation of this framework is the ``Area Plot'', which visualizes the trade-off between solution quality and cumulative runtime especially for hybrid quantum computing algorithms like QAOA and Quantum Annealing \cite{lubinski2024optimization}. Herein,
\begin{itemize}
    \item \textit{Color} encodes the solution quality (e.g., Approximation Ratio);
    \item \textit{Width} represents the execution time per iteration;
    \item \textit{X-axis} corresponds to cumulative runtime.
    \item \textit{Y-axis} indicates the problem size.
\end{itemize}
\end{enumerate}
For optimization heuristics such as QAOA and quantum annealing, the Advanced QED-C framework enables systematic analysis of the trade-off between solution quality and computational cost. In particular, it allows researchers to identify the \textbf{performance crossover point} at which a quantum solver achieves a lower \textit{Time-to-Solution} ($TTS$) than state-of-the-art classical optimization algorithms. Consequently, the framework provides an important methodology for assessing the practical potential of quantum optimization techniques and for identifying problem regimes where quantum approaches may offer computational advantages.

\subsubsection{TAQOS: Tight Analysis of Quantum Optimization Systems}
TAQOS provides a benchmarking protocol designed to evaluate the performance of quantum optimization systems in the NISQ era \cite{Gilbert2023}. The framework enables fair comparison between quantum and classical optimization algorithms by considering multiple performance dimensions. Unlike conventional benchmarks that focus mainly on solution quality, TAQOS evaluates optimization systems using a combination of solution quality, computational time, energy consumption, and robustness across different problem instances.

Let $P=\{P_1,P_2,...,P_n\}$ denote a set of combinatorial optimization problems and ${\cal I}$ represent the set of possible instances associated with a given problem. For an instance $I_i \in {\cal I}$, the quality of a solution $s_i$ is evaluated using an objective function $c(s_i)$. Considering a reference solution with cost $c_{ref}$, the normalized solution quality ratio is defined as
\begin{equation}
r_{\mathrm{ref}}(c(s_i), c_\mathrm{ref}) = \frac{c_\mathrm{ref} - c(s_i)}{c_\mathrm{ref}} .
\end{equation}

\noindent This ratio allows consistent comparison of solution quality across different problem instances. A negative value of $r_{\mathrm{ref}}$ indicates that the obtained solution is better than the reference solution.

TAQOS also evaluates how frequently an algorithm produces near-optimal solutions across a set of instances. This is measured by the fraction of instances that achieve a solution within a tolerance $\varepsilon$ of the reference solution
\begin{equation}
r_\varepsilon({\cal I}) =
\frac{|\{ I_i \in {\cal I} \mid r_{\mathrm{ref}}(c_q^*, c_\mathrm{ref}) < \varepsilon \}|}{|{\cal I}|}.
\end{equation}

\noindent Here $\varepsilon$ typically represents a small deviation from optimality, such as $1\%$, $5\%$, or $10\%$.
In addition to solution quality, TAQOS considers the full wall-clock time required to solve the optimization problem, including both quantum and classical processing stages such as problem mapping, circuit execution, parameter optimization, and post-processing. Energy consumption is also identified as an important benchmarking dimension, although precise measurements remain difficult for current quantum hardware.
Another key component of the TAQOS framework is instance coverage analysis. This metric evaluates whether the selected benchmark instances adequately represent the diversity of the problem space. For a normalized instance feature $f(I_i)$, the coverage interval is defined as
\begin{equation}
c_\varepsilon(f, I_i) = [f(I_i) - \varepsilon, f(I_i) + \varepsilon] \cap [0,1].
\end{equation}
\noindent The overall coverage of the benchmark instance set is
\begin{equation}
C_\varepsilon(f, I) =
\bigcup_{I_i \in I} c_\varepsilon(f, I_i).
\end{equation}
This analysis helps ensure that benchmarking results are not biased toward a narrow subset of favorable problem instances. The TAQOS protocol has been demonstrated on combinatorial optimization problems such as Max-Cut to illustrate how the framework enables systematic comparison between quantum and classical heuristics. By integrating solution quality, runtime analysis, and instance diversity, TAQOS provides a reproducible methodology for evaluating quantum optimization systems.

The significance of TAQOS arises from the following features:

\begin{itemize}
    \item \textbf{Multi-dimensional benchmarking:} evaluates solution quality, runtime, and energy considerations simultaneously.
    \item \textbf{Reproducibility:} provides a structured protocol for consistent comparison between quantum and classical optimization algorithms.
    \item \textbf{Instance diversity:} ensures that benchmarking experiments include a representative set of problem instances.
    \item \textbf{Transparent performance reporting:} accounts for both classical and quantum computational overheads.
\end{itemize}

Although these benchmarking frameworks share the common objective of evaluating quantum computing performance, they address different layers of the computational stack. QASMBench focuses primarily on circuit-level characteristics and hardware behaviour, enabling detailed analysis of how quantum circuits interact with NISQ devices. In contrast, QUARK evaluates the end-to-end performance of quantum applications by considering the entire workflow from problem formulation to solution validation. The Advanced QED-C framework places particular emphasis on the trade-off between runtime and solution quality, which is critical for real-world optimization tasks where computational efficiency directly impacts practical applicability. TAQOS further extends benchmarking methodology by incorporating multidimensional evaluation criteria, including solution quality, computational time, energy considerations, and instance diversity. Taken together, these frameworks provide complementary perspectives for assessing quantum optimization algorithms across hardware, algorithmic, and application levels. Among recently proposed approaches, the benchmarking framework introduced by Sharma and Lau \cite{Sharma2025comparative} provides a systematic methodology for evaluating quantum optimization algorithms on challenging combinatorial problems such as MDKP, MIS, QAP, and MSP. By comparing algorithms including QAOA, VQE, and qubit-efficient encoding strategies across multiple performance metrics, the framework helps identify the practical strengths and limitations of near-term quantum optimization techniques. On the other hand, Bucher et al. \cite{Bucher2024} introduce a set of methodological guidelines for conducting fair and reproducible benchmarking of quantum optimization algorithms against classical solvers. Their work emphasizes consistent problem formulations, realistic runtime accounting, and standardized evaluation metrics, thereby enabling more reliable assessment of potential quantum advantage in combinatorial optimization. The comparison between these benchmarking frameworks has been summarised in Table \ref{tab:bench_compare}.

\begin{table}[h]
\tiny
\centering
\caption{A Comparison of Selected Benchmarking Frameworks}
\label{tab:bench_compare}
\begin{tabular}{|p{0.10\textwidth}|p{0.15\textwidth}|p{0.15\textwidth}|p{0.15\textwidth}|p{0.15\textwidth}|p{0.15\textwidth}|}
\hline
\textbf{Benchmarking Framework} & \textbf{Core Goal} & \textbf{Level of Analysis} & \textbf{Algorithms} & \textbf{Problem Class} & \textbf{Industry Domains} \\
\hline
QASMBench \cite{Ang2023} & A low-level benchmark suite of diverse quantum circuits (kernels) written in OpenQASM. & Hardware/system-level; gate-level circuit analysis & Quantum circuit kernels (e.g., QFT, Grover, arithmetic circuits, VQE circuits) & Quantum algorithm kernels across multiple domains (chemistry, arithmetic, ML, optimization) & Hardware evaluation, compiler benchmarking \\
\hline
QUARK \cite{Rudi2022, kiwit2024benchmarking} & Application-centric benchmarking to compare end-to-end performance of quantum, hybrid, and classical solvers on specific problems. & Application-level & QAOA, quantum annealing, quantum machine learning (e.g., QGANs, QCBMs) & Robot path planning, vehicle configuration optimization, and machine learning tasks & Logistics (e.g., robot path optimization, auto-carrier loading), manufacturing, finance \\
\hline
Advanced QED-C \cite{lubinski2024optimization} & Application-oriented benchmarking for evaluating practical quantum utility by analyzing solution quality and time-to-solution trade-offs. & Application-level; hybrid quantum–classical benchmarking & QAOA, quantum annealing & Combinatorial optimization workloads (e.g., Max-Cut and related optimization problems) & Portfolio optimization, industrial scheduling, transportation, logistics \\
\hline
TAQOS \cite{Gilbert2023} & Provides a ``Tight Analysis'' protocol for quantum optimization solvers using a multi-metric evaluation framework. & Application-level; multi-dimensional benchmarking & QAOA, adiabatic quantum optimization (AQO) & Combinatorial optimization problems (e.g., Max-Cut) & Logistics, finance, network design, and related optimization domains \\
\hline
Sharma \& Lau \cite{Sharma2025comparative} & A framework to systematically evaluate and compare different quantum optimization techniques. & Application benchmarking through practical implementations & VQE, QAOA & Multi-dimensional knapsack problem (MDKP), maximum independent set (MIS), quadratic assignment problem (QAP), market share problem (MSP) & Supply chain, logistics, finance, telecommunications and networking \\
\hline
Bucher et al. \cite{Bucher2024} & Provides methodological guidelines to establish fair benchmarking comparisons between quantum and classical optimization solvers. & Application benchmarking at the methodological level & Quantum annealing, QAOA & Combinatorial optimization problems (e.g., Max-Cut, TSP) & VLSI design, logistics, finance \\
\hline
\end{tabular}
\end{table}

\section{Industry Adoption and Broader Relevance}
\label{sec:Industry}

\subsection{Bridging Problem Classes to Industry Relevance}
Industrial use cases can be viewed as practical instances of the abstract problem classes commonly used in benchmarking studies and optimisation libraries. Relating these mathematical formulations to real-world terminology helps clarify how quantum optimisation research connects to operational problems. In combinatorial optimisation, this typically involves mapping standard formulations (e.g., knapsack, Max-Cut, market split) to decision-making tasks encountered in applied settings. For example, financial institutions do not describe their task as solving a QUBO instance, but rather as optimizing portfolios under risk–return trade-offs. The underlying formulation, however, is often reduced to a QUBO problem. Similarly, transportation planners refer to minimizing travel time, fuel consumption, or delivery cost rather than explicitly solving a Vehicle Routing Problem (VRP). In the energy sector, operators focus on efficient distribution and grid stability rather than framing their objective as a Network Design optimization problem.

In this work, we reference the problem classes identified in major optimization benchmarks such as QOBLIB \cite{koch2025quantum}, the SMU repository \cite{Sharma2025comparative}, and HamLib \cite{Sawaya2024HamLib}, and map them to their industrial counterparts. The discussion focuses on industries with significant reliance on combinatorial optimization: Finance, Telecommunications, Energy, Logistics, and Transportation. Most of these problem classes are NP-Hard or NP-Complete, highlighting the computational limitations of classical approaches and motivating the exploration of quantum advantage.

For clarity, we group the selected problems into two categories: Resource Distribution and Graph Optimization.

\paragraph{Resource Distribution}

Resource distribution problems can be broadly divided into partition and connectivity classes.

A \textbf{partition problem} divides a system into distinct, non-overlapping subsets to optimize balance or separation. The \textit{Low Autocorrelation Binary Sequence} (LABS) problem seeks binary sequences with minimal autocorrelation between shifted copies, thereby reducing interference in radar, Wi-Fi, and GPS applications \cite{ukil2010low}. The \textit{Maximum Independent Set} (MIS) selects the largest subset of mutually non-adjacent elements, while \textit{Max-Cut} partitions a graph to maximize the weight of edges between subsets. Ref.~\cite{wu2024} demonstrates a hybrid formulation where MIS derived from an Ising model is solved using Max-Cut solvers, enabling improved scalability through transformation into a more mature algorithmic framework.

\textbf{Connectivity problems} focus on designing cost-efficient and resilient network structures. \textit{Steiner Tree Packing}, considered a subclass of \textit{Network Design}, identifies minimal-cost sub-networks connecting key nodes. Applications include optimal heating systems and optical fibre distribution \cite{braunstein2018cavity}.

\paragraph{Graph Optimization}

Graph optimization problems address the assignment, routing, or allocation of tasks and resources within networks. These are divided into Routing/Scheduling, Assignment, and Allocation classes.

\textbf{Routing/Scheduling} problems optimize order and traversal. The \textit{Vehicle Routing Problem} (VRP) generalizes the \textit{Travelling Salesman Problem} (TSP) to multiple vehicles and capacity constraints \cite{Toth2014}. Sports Tournament Scheduling assigns matches to venues and time slots while minimizing travel and ensuring fairness \cite{KENDALL20101}.

\textbf{Assignment problems} map entities to locations or configurations. The \textit{Quadratic Assignment Problem} (QAP) accounts for pairwise interaction costs between assigned elements \cite{Tjalling1957}. It applies to facility layout and logistics planning, and TSP can be interpreted as a special case of QAP \cite{verel2024}. The \textit{Minimum Birkhoff Decomposition} is used in high-speed networking to decompose traffic matrices into feasible switching schedules \cite{valls2020}.

\textbf{Allocation problems} determine optimal distribution of limited resources. \textit{Portfolio Optimization} balances return and risk under constraints \cite{VALLE2025106918}. \textit{Bin Packing} minimizes container usage \cite{Coffman2012}, while the \textit{Multi-Dimensional Knapsack Problem} extends this to multiple capacity constraints \cite{Kellerer2004}.

\begin{figure}
\centering
\includegraphics[width=0.9\linewidth]{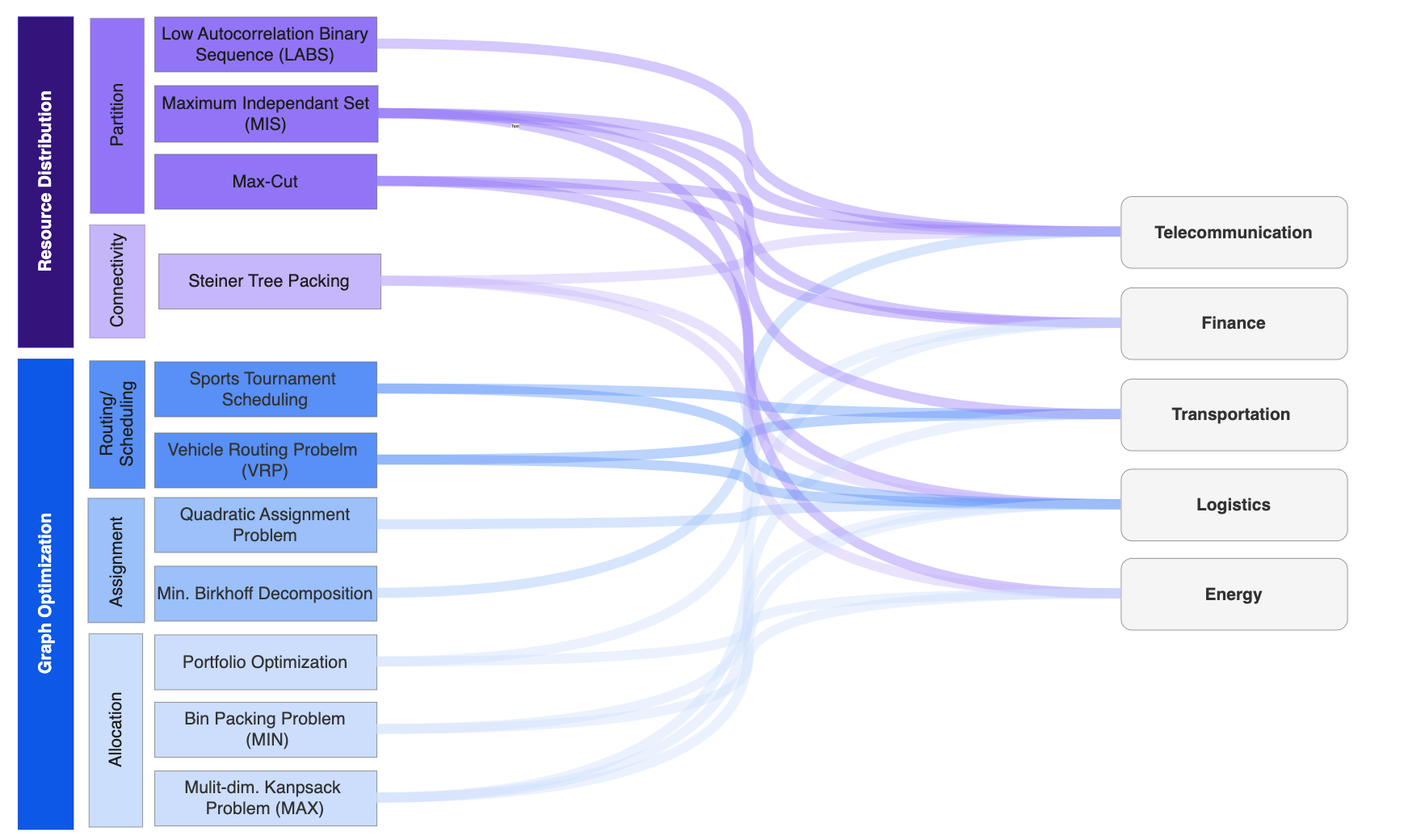}
\caption{Mapping of representative combinatorial optimisation problem classes to their corresponding industry sectors. The problem classes are divided into Resource Distribution and Graph Optimization, supporting applications across telecommunications, finance, transportation, logistics, and energy.}
\label{fig:IndustryDiag}
\end{figure}

\subsection{Technology Readiness Level of Hybrid Algorithms}

The Technology Readiness Level (TRL) framework, introduced by NASA \cite{Mankins1995}, quantifies technological maturity. It has since become a standard metric across industries evaluating emerging technologies. In the quantum domain, TRL assessments are increasingly used alongside benchmarking frameworks such as TAQOS and QED-C.

TRL 1–3 denote conceptual and proof-of-concept stages; TRL 4–6 represent validation in relevant environments; TRL 7–9 indicate system-level maturity and operational deployment.

\paragraph{Direct Cost Minimization}

This category includes quantum annealing and QAOA. These approaches encode objective functions directly into energy minimization problems, applicable to Max-Cut, MIS, QAP, and Steiner Tree. Ref.~\cite{mesman2022} introduces QPack and defines the Quantum Technology Readiness Level (QTRL). Current QAOA implementations typically fall within QTRL 3–5, corresponding to experimental validation without industrial deployment. In contrast, quantum annealing has reached TRL 7–9, supported by documented industrial applications \cite{Yarkoni2022, Mappas2025}.

\paragraph{Return Maximization}

Quantum Reinforcement Learning (QRL) remains largely theoretical, with maturity around QTRL 2–4 \cite{kaldari2025}. Real-time environment interaction and scalability constraints limit deployment readiness. Nonetheless, exploratory applications include last-mile delivery \cite{moosavi2025}, financial sector rotation \cite{chen2025}, and antenna path design \cite{Arani2024, tomar2025}.

\paragraph{Likelihood Maximization}

Quantum generative optimization maximizes likelihood functions over parametrized quantum states. Applications include stochastic portfolio optimization using S\&P~500 data \cite{zoufal2019, Alcazar2024} and molecular structure optimization \cite{Kao2023}. Despite promising demonstrations, implementations remain at QTRL 2–4.

\subsection{From Benchmarking to Industry}

Optimization serves as a primary interface between quantum theory and practical deployment. While theoretical quantum advantage remains a central research goal, industrial adoption requires demonstrable performance, scalability, and reliability.

A significant gap persists between theoretical advances and industrial integration. Rigorous benchmarking therefore becomes a strategic necessity. Benchmarking frameworks provide structured evaluation methodologies that link theoretical developments to industrial applicability.

The critical roles of benchmarking include:

\begin{itemize}

\item \textbf{Securing Investment:} Benchmarking provides empirical evidence necessary for investment decisions in costly quantum hardware.

\item \textbf{Standardization:} Establishes uniform metrics enabling fair cross-vendor comparison.

\item \textbf{Performance–Utility Trade-off:} Evaluates solution quality, time-to-solution, and cost-to-solution beyond hardware fidelity.

\item \textbf{Fair NISQ Comparison:} QASMBench demonstrated fidelity differences across IBM-Q systems \cite{Ang2023}. QUARK differentiated D-Wave 2000Q and Advantage 4.1 \cite{Rudi2022}.

\item \textbf{Cross-Vendor Comparison:} Benchmarking compares IBM-Q, IonQ, Rigetti, and D-Wave systems. For instance, advanced QED-C highlights trade-offs across hardware families, summarized in Table~\ref{tab:platforms_QEDC}.

\item \textbf{Understanding Quantum Advantage:} Ref.~\cite{Bucher2024} shows classical heuristics dominate small instances, while quantum solvers exhibit potential scaling benefits.

\item \textbf{Establishing Realistic Expectations:} Benchmarking identifies conditions where quantum methods underperform, preventing unrealistic hype. A comparative study of major benchmarking frameworks is provided in Table~\ref{tab:bench_compare}.

\end{itemize}

\begin{table}[h]
\tiny
\centering
\caption{Comparative performance characteristics across hardware families.}
\label{tab:platforms_QEDC}
\begin{tabular}{|p{0.25\textwidth}|p{0.25\textwidth}|p{0.20\textwidth}|p{0.20\textwidth}|}
\hline
\textbf{Platform Type} & \textbf{Characteristic Advantage} & \textbf{Limiting Factor} & \textbf{Observed Scaling Trend} \\
\hline
Ion Trap (IonQ Aria) & High fidelity, all-to-all connectivity & Long per-shot latency & Execution time $\propto$ number of shots \\
\hline
Superconducting (IBM Guadalupe) & High throughput, short per-shot latency & Connectivity constraints, SWAP overhead & Weak dependence on shot count \\
\hline
Quantum Annealer (D-Wave Advantage) & Fast single-step evolution & Embedding overhead & Anneal time scales mildly with problem size \\
\hline
\end{tabular}
\end{table}

\section{Conclusion}\label{sec:conclusion}
To conclude, this work bridges quantum optimization theory and its practical use in industry. It shows where existing methods fall short when applied to real-world problems. Optimization problems represent one of the most natural and accessible initial domains for quantum computing as it approaches commercial feasibility. Optimization plays a central role across virtually all industries, from financial portfolio management and healthcare resource allocation to energy grid configuration. A key limitation of classical optimization methods is their difficulty in scaling efficiently as problem complexity increases, particularly for problems classified as NP-Hard or NP-Complete. Quantum computing offers the prospect of a potential “quantum advantage,” enabling alternative exploration strategies of the solution space through inherently quantum mechanical effects, which may help mitigate issues such as trapping in local minima that frequently affect classical heuristics. This review has analyzed the current landscape of quantum optimization by examining algorithmic paradigms, benchmarking frameworks, and application-driven problem formulations that are most relevant for near-term quantum technologies. \\

A clear dichotomy in operational maturity emerges from the current landscape of quantum optimization algorithms. For combinatorial optimization problems mapped to Ising/QUBO formulations, quantum annealing exhibits the highest level of technological readiness (TRL 7–9), with established hardware implementations and early industrial deployments. In contrast, gate-based approaches, most notably QAOA, are currently positioned at QTRL 3–5, demonstrating significant experimental validation but not yet achieving full industrial readiness. More recent variational paradigms, such as QRL and QGM, remain largely at the theoretical and simulation stages (QTRL 2–4), with practical implementations still in their infancy. \\
The transition from theoretical promise to validated, application-oriented utility critically depends on rigorous benchmarking. Beyond hardware-level performance indicators, advanced evaluation frameworks such as QUARK, the Advanced QED-C Framework, and TAQOS emphasize industry-relevant metrics, including Time-to-Solution (TTS), solution Quality (Q), and Validity (V). Such benchmarking efforts are essential to bridge abstract problem classes — including Max-Cut, Vehicle Routing Problem, and Portfolio Optimization — with concrete implementations in strategically relevant industrial sectors. \\
Despite tangible progress, substantial challenges remain. Variational algorithms, for instance, face the theoretical obstacle of barren plateaus, where gradients required for parameter optimization vanish exponentially, severely limiting trainability. In the long term, the development of fault-tolerant quantum computers (FTQCs) remains a central objective. In the short to medium term, however, progress will likely depend on application-specific co-design strategies, in which hardware architectures, algorithmic structures, and problem encodings are jointly optimized. This integrated approach raises a fundamental open question: the identification of optimization tasks that are genuinely “quantum-native” while remaining classically intractable. In other words, the field must determine which problems align most naturally with the physical capabilities and constraints of near-term quantum devices. Although several industry-relevant “intractable decathlon” problems have been discussed, the systematic identification of such tasks remains an open challenge for the community. \\
In summary, this review has framed quantum optimization not merely as a theoretical aspiration, but as an emerging technological domain supported by concrete experimental advances. While transformative large-scale applications have not yet been fully realized, the field has clearly moved beyond purely speculative discussions. Sustained progress will require the simultaneous advancement of hardware scalability, algorithmic innovation, and robust domain-specific implementation frameworks.

\section*{References}
\bibliographystyle{unsrt}
\bibliography{References} 

\end{document}